\documentclass[twocolumn,twoside,slac_two]{revtex4}
\usepackage{graphicx}
\usepackage{fancyhdr}
\begin{document}

\newcommand{\aj}{AJ}
\newcommand{\apjs}{ApJS}
\newcommand{\apjl}{ApJ}
\newcommand{\mnras}{MNRAS}
\newcommand{\pasj}{PASJ}
\newcommand{\aap}{A\&A}
\newcommand{\na}{New Astr.}
\newcommand{\nar}{New Astr. Rev.}
\newcommand{\araa}{ARA\&A}
\newcommand{\apss}{Ap\&SS}
\newcommand{\jcap}{JCAP}

\def\be{\begin{equation}}
\def\ee{\end{equation}}
\def\ba{\begin{eqnarray}}
\def\ea{\end{eqnarray}}
\def\d{\delta}
\def\e{\epsilon}
\def\f{\varphi}
\def\k{\varkappa}
\def\tde{\tilde}
\def\p{\partial}
\def\ms{\mathstrut}
\def\s{\strut}
\def\ds{\displaystyle}
\def\ts{\textstyle}
\def\b{\boldsymbol}
\def\r{\mathrm}
\def\G{\Gamma}
\def\sun{\odot}
\def\c454{3C454.3}
\def\ag{{\it AGILE}}
\def\fer{{\it Fermi}/LAT}
\def\pks{{PKS~2155$-$304}}
\def\opks{{PKS~1510$-$089}}
\def\mkn501{{Mkn~501}}

\title{Star-Jet Interactions and Gamma-Ray Outbursts from  3c454.3}

%


\author{M.V.~Barkov$^{1,2,3}$,
D.V.~Khangulyan$^{4}$,  
V.~Bosch-Ramon$^{5}$,
F.A.~Aharonian$^{6,2}$,
A.V.~Dorodnitsyn$^{7,8}$
}
\affiliation{$^{1}$Astrophysical Big Bang Laboratory, RIKEN, 2-1 Hirosawa, Wako, Saitama 351-0198, Japan}
\affiliation{$^{2}$Max-Planck-Institut f\"ur Kernphysik,
Saupfercheckweg 1, D-69117 Heidelberg, Germany}
\affiliation{$^{3}$Space Research Institute RAS, 84/32 Profsoyuznaya Street, Moscow, 117997, Russia}
\affiliation{$^{4}$Institute of Space and Astronautical Science/JAXA, 3-1-1 Yoshinodai, Chuo-ku, Sagamihara, Kanagawa 252-5210, Japan}
\affiliation{$^{5}$Departament d'Astronomia i Meteorologia, Institut de Ci\`encies del Cosmos (ICC), Universitat de Barcelona (IEEC-UB),
Mart\'i i Franqu\`es, 1, E08028, Barcelona, Spain}
\affiliation{$^{6}$Dublin Institute for Advanced Studies, 31
Fitzwilliam Place, Dublin 2, Ireland}
\affiliation{$^{7}$Laboratory for High Energy Astrophysics, NASA Goddard Space Flight Center, Code 662, Greenbelt, MD, 20771, USA}
\affiliation{$^{8}$Department of Astronomy/CRESST, University of Maryland, College Park, MD 20742, USA}




\begin{abstract}
{We propose a model to explain the ultra-bright GeV gamma-ray flares observed from  the blazar \c454. The model is based on the
concept of a relativistic jet interacting with  compact gas  condensations produced when a star (red giant)  crosses the jet close to
the central black hole.  The study   includes an analytical treatment of the evolution of the envelop lost by the star
within the jet, and calculations of the related  high-energy radiation \cite{kbbad13}.
The model readily explains the day-long, variable on timescales of hours, GeV gamma-ray flare from \c454,
observed during November 2010 on top of a weeks-long plateau. In the proposed scenario,  the plateau state 
is caused by a strong wind generated by the heating of the star atmosphere by nonthermal particles accelerated at the jet-star interaction
region. The flare itself could be  produced by a few clouds of matter lost by the red giant after the initial impact
of the jet. In the framework of the proposed scenario, the observations   constrain the  key model parameters of the source, including  the 
mass of the central black hole: $M_{\rm BH}\simeq 10^9 M_{\odot}$, the total jet power: $L_{\rm j}\simeq 10^{48}\,\rm erg\,s^{-1}$,
and the Doppler factor of  the gamma-ray emitting clouds, $\delta\simeq 20$. Whereas we do not specify the particle acceleration mechanisms, 
the potential  gamma-ray production processes  are discussed and compared in the context of the proposed model. We argue that
synchrotron radiation of protons has certain advantages compared to other radiation channels of directly accelerated
electrons.}
\end{abstract}

\maketitle

\thispagestyle{fancy}

\section{INTRODUCTION}
\c454 is a powerful flat-spectrum radio quasar located at a redshift $z_{\rm rs}=0.859$. This source is 
very bright  in the GeV
energy range; during strong flares, its  apparent (isotropic) luminosity can reach $L_{\gamma}\gtrsim 10^{50}$~erg~s$^{-1}$
\cite[e.g.][]{agile10,fermi10,agile11,fermi11_3C}. The mass of the central black hole (BH) in
 \c454 is estimated in the range $M_{\rm BH}\approx (0.5-4)\times 10^9 \,M_{\odot}$ \cite{gcj01,bgf11}. This implies an Eddington luminosity
$L_{\rm Edd}\approx (0.6-5)\times 10^{47}\mbox{erg s}^{-1}$, which is several orders of magnitude below $L_\gamma$.
Although the  large  gap  within  $L_{\rm Edd}$ and $L_\gamma$  is naturally explained by  
relativistic Doppler  boosting,  the estimates of the jet power during these flares appear,
in any realistic scenario, close to or  even larger than  the  
Eddington luminosity \cite[][]{bgf11}.   

The   GeV emission from \c454 is highly erratic, with variability  timescales as short as 3~hr, 
as reported, in particular,  for
the  December 2009 flare  \cite{fermi10}. The most  spectacular  flare regarding both 
variability and gamma-ray
luminosity was  observed in November 2010 by  \ag and \fer \cite{agile11,fermi11_3C} telescopes. 
During this high state, with the most active
phase lasting for 5 days, the apparent  luminosity in GeV  achieved  $L_{\gamma}\approx 2\times10^{50} \mbox{
erg s}^{-1}$. Around the flare maximum, the rising time was $t_{\rm
r}\approx 4.5$~hr, and the decay time, $t_{\rm f}\approx 15$~hr. The detection of photons with energies up to $\approx
30$~GeV, the short variability, and the contemporaneous X-ray flux constrain the Doppler boosting of the emitter to
$\delta_{\rm min}\gtrsim 16$ to avoid severe internal $\gamma\gamma$ absorption in the
X-ray radiation field \cite{fermi11_3C}.

A remarkable feature of the gamma-ray emission from \c454 is the so-called plateau phase 
revealed  during the bright flare  in 2010.    It is characterized by a long-term 
brightening of the source,   a  few weeks before the appearance of the main flare. Such  plateau states
have been  observed by \fer for  three flares
\cite[e.g.][]{fermi10,fermi11_3C}, with the plateau emission being about an order of magnitude  
fainter than that of the main flare.

Remarkably,  the rapid gamma-ray variability of \c454  is accompanied by an activity at lower energies. The simultaneous 
multiwavelength observations  of the source during flares  have revealed a strong correlation with optical and X-rays. It has been
interpreted as evidence that the gamma-ray source is located upstream from the core of the 43~GHz radio source, 
which is at a distance $z<$~few~pc from the central BH \cite[see, e.g.,][]{jml10,jor12,wmj12}. 

Over the recent years,  several  works have attempted to explain the flaring gamma-ray activity of \c454 
within the framework of the standard synchrotron self-Compton (SSC)
or external inverse-Compton (EIC) models \cite{kg07,gft07,smm08,bgf11}. 
In the SSC scenario,  it is possible to  reproduce
the spectral energy distribution (SED) from optical wavelengths to  gamma-rays. 
In these models  most of the jet power is (unavoidably) carried by protons,
and only a small fraction is contained  in relativistic electrons and the magnetic field.  
The required proton-to-Poynting flux
ratio of $L_{\rm p}/L_{\rm B}\sim 100$ is quite large. Such a  
configuration would be hard to reconcile, at least in the gamma-ray emitting region 
close to the central BH, with an undisturbed jet which is launched by the Blandford-Znajek  \cite{BZ77} type
(BZ) process, 
in which the luminosity of the jet is dominated by Poynting flux and the jet
consists of $e^\pm$-pairs. 
In this regard we should mention that recent relativistic magnetohydrodynamical (MHD) simulations of 
jet acceleration yield much less efficiency of  conversion of the  
magnetic energy into bulk motion kinetic  energy; these calculations  
\cite[][]{kbvk07,kvkb09} predict a quite modest ratio  $(L_{\rm p}+L_{e^\pm})/L_{\rm B}\lesssim 4$.

The  jet-RG interaction (JRGI) scenario has been invoked 
to explain the day-scale flares in the  nearby
non-blazar type AGN  M87 \cite{bab10,bpb12},  
It  has been  applied also to the TeV blazar \pks \  \cite{babkk10} to demonstrate that 
the jet-driven acceleration of debris from the  RG atmosphere can explain ultra-fast variability of 
very high-energy gamma-ray emission  on  timescales as short as $\tau\sim 200$~s. 
A  distinct   feature of the JRGI scenario is  the  high magnetization ($L_{\rm B}/L_{\rm p,e}\gg  1$) of the relativistic 
flows located at sub-parsec distances, where the gamma-ray production 
supposedly takes place.  Although  the  strong  magnetic field, $B \geq 10$~G,  dramatically reduces the efficiency of  the inverse 
Compton scattering of electrons, it opens an alternative channel of gamma-ray production through synchrotron radiation of protons
\cite{ah00,mp01}. 
The latter can be effectively realized only in the case of acceleration of protons to the highest possible energies, 
up to $10^{20}$eV.  
Thus the second (somewhat ``hidden")   requirement of this model is a very effective acceleration of protons with a rate close to the
theoretical limit dictated by the classical electrodynamics \cite{a02}.  

It is interesting to note that also inverse Compton  models can be 
accommodated, at least in principle, in the JRGI  scenario. Moreover, unlike most of the leptonic models 
of powerful blazars, in which the  requirement of a very low magnetic field, 
implying a deviation from the equipartition condition by orders of magnitude, generally 
is not addressed and explained, the JRGI scenario  can offer
a natural way for leptonic models to be effective assuming that  the gamma-ray  emission 
is produced through the inverse Compton scattering in shocked clouds originally weakly   
magnetized \cite[see ][]{bba12}. 

In this work, we show that  the JRGI scenario gives a viable mechanism for the explanation of the flares
seen in \c454. 
We also  argue  that  within this model   the plateau state can form due
to the interaction of the jet with a stellar wind excited by 
nonthermal (accelerated)  particles that penetrate into the red giant atmosphere.

\begin{figure}
\includegraphics[width=0.47\textwidth,angle=0]{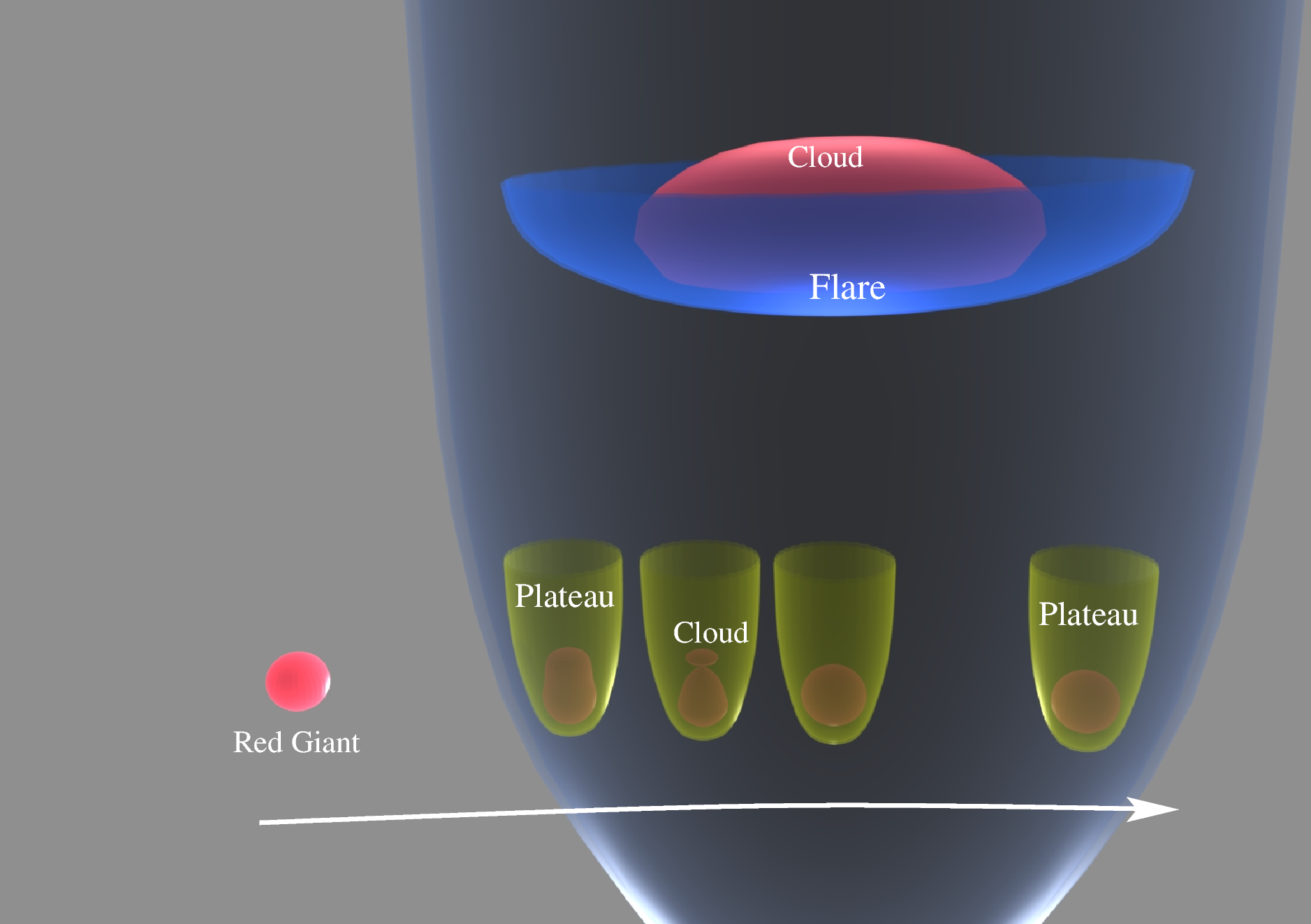}
\caption{Sketch for the JRGI scenario, in which a star moving from left to right penetrates into the jet. 
The star external layers are shocked and carried away, and a cometary tail, origin of the plateu emission, forms. 
The acceleration and expansion of the bigger clouds from the initially blown-up external layers of the star would 
lead to the main flare.}
\label{sketch}
\end{figure}

\section{Star-Jet interaction Scenario}
\label{sec:scenario}

In the {\it fast cooling regime}, the proper intensity of  the nonthermal emission, i.e. the intensity 
in the blob co-moving reference frame, is  proportional to the energy released at the jet-blob interface. This energy release can be 
characterized by a simple dynamical model, which describes the acceleration of the blob by the jet ram pressure. In this model
there are just a few relevant parameters that describe the basic properties of the jet and the blob: the jet ram pressure ($P_{\rm j}$) and 
bulk Lorentz factor ($\Gamma_{\rm j}$), and the blob mass ($M_{\rm b}$) and  radius ($r_{\rm b}$; or, equivalently, its cross-section: $S_{\rm b}=\pi r_{\rm b}^2$) 
\cite[for details, see][]{babkk10}. 
The time dependence of the intensity of the jet/blob interaction
corrected for the Doppler boosting can be treated as a first-order approximation for the radiation
lightcurve.

The mass of the cloud $\Delta M$ formed at the initial stage can be estimated by comparing the jet ram pressure, 
\begin{equation}
P_0\simeq \frac{L_{\rm j}}{c\pi\omega_0^2}\,
\label{eq_ram}
\end{equation}
with the gravitational force.  
Here, $\omega_0$ is the jet cross-section at the star crossing height. This gives  the
following estimate:
\begin{equation}
\Delta M=\frac{\pi P_0 R_{*}^4}{G M_{*}} 
\label{dms}
\end{equation}
where $M_{*}$ and $R_*$ are the RG mass and radius, respectively.

Since the initial size of the expelled cloud should be comparable to the size of the star, it is possible to estimate the cloud
expansion time as $t_{\rm exp}\propto 2R_*/c_{\rm s}$, where $c_{\rm s}$ is the sound speed of the shocked material:
$c_{\rm s}\approx\left[(4\pi R_*^3/3)\gamma_{\rm g} P_0/M_{\rm c}\right]^{1/2}$. 
The cloud expansion time is 
\begin{equation}
t_{\rm exp} \approx A_{\rm exp} 
\left(\frac{M_{\rm c}}{\gamma_{\rm g} R_{*} P_0}\right)^{1/2},
\label{eq_expansion}
\end{equation}
where $\gamma_{\rm g}=4/3$ is the plasma adiabatic coefficient, and according to the RHD simulation by \cite{bpb12}, a value of $1.5$ can be adopted for $A_{\rm 
exp}$.

The blob acceleration occurs on a timescale of \cite{babkk10} 
\begin{equation}
t_{\rm acc}\approx\left\{{\frac{z_0}{c}\quad {\rm if}\quad D<1\atop\,\frac{z_0}{c}\frac{1}{D}\quad {\rm if}\quad D>1\,.}\right.
\label{dtpeak}
\end{equation}
The $D$--parameter that will be often used in the paper has a simple meaning. It is a dimensionless inverse mass of the blob: 
\begin{equation} 
D\equiv {P_{\rm 0}\pi r_{\rm b}^2z_0\over4c^2M_{\rm b}\Gamma_0^3}
\,. 
\label{DD} 
\end{equation} 
The above timescale corresponds to the blob acceleration in the laboratory reference frame. However, since the blob gets
accelerated towards the observer, the emission delay, as seen by the observer, should be approximately corrected by a 
factor of
$1/(2\Gamma_0^2)$. Thus, the observed peak of the emission should be delayed by a time interval of
\begin{equation}
\Delta t=t_{\rm exp}+t_{\rm acc}/(2\Gamma_0^2)\,.
\label{eq_delay}
\end{equation}

The emission produced by lighter clouds allows an estimation of the time
required for the star to cross the jet. Once the star enters into the jet, the process of jet-star interaction should
proceed steadily, with the production of these lighter clouds being roughly constant on average. 
Thus, the whole duration of the light cloud-associated 
emission, if observed, can be taken as a direct measurement of the jet crossing time $t_0\approx 2\omega_0/V_{\rm orb}$,
where $V_{\rm orb}\lesssim\sqrt{2GM_{\rm BH}/z_0}$ is the star velocity. 
Adopting the paradigm of magnetically-accelerated jets, it is 
possible
to derive a very simple expression for this timescale:
\begin{equation}
t_0\gtrsim 2^{3/2} z_0/c\,.
\label{eq_cross_time}
\end{equation}
In this way, the duration of the jet-star interaction is determined by the interaction distance from the central BH.

Assuming a fixed efficiency $\xi$ in the blob reference frame
for the transfer of jet power to nonthermal gamma rays (where $\xi\ll 1$), and correcting for Doppler boosting, one can 
estimate the luminosity of a blob:
\begin{equation}
L_{\gamma}=4\xi c F_{\rm e} P_0 \Gamma_0^2 \pi r_{\rm b}^2\,,
\label{eq_lum}
\end{equation}
where the correction function $F_{\rm e}$ depends on time; or, equivalently, on the blob location in the jet. We note that the structure of the jet, i.e., 
the dependence of the jet Lorentz factor on $z$, determines the actual dependence of $F_{\rm e}$ on $z$.

The maximum value of $F_{\rm e}$ monotonically depends on the
$D$ parameter, approaching a value of $0.4$ if $D\gtrsim 1$ and being
$\sim 0.1$ for $D=0.1$. This relatively weak dependence allows us to derive the
maximum blob luminosity. Also it is possible to obtain
an  estimate of the total energy emitted by a blob 
or  an ensemble of sub-blobs as a result of the fragmentation of the original cloud ($M_{\rm c}=\sum M_{\rm b}$), 
\begin{equation}
E_{\gamma}\simeq 8\xi \bar{F_{\rm e}} M_{\rm b/c}c^2\Gamma_0^3\,,
\label{eq_toten}
\end{equation}
which accounts for the total energy transferred by the jet to a blob during the
acceleration process, $M_{\rm b}c^2\Gamma_0$, and for the anisotropy of
the emission due to relativistic effects represented by the  factor 
$\Gamma_0^2$.

\section{The November 2010 Flare}

\subsection{General structure of the active phase}\label{3c}

The total apparent energy of the GeV gamma-ray radiation detected during the flare observed from \c454 in November 2010 was about
$E_{\rm tot}\approx L_{\gamma}\Delta t/(1+z_{\rm rs}) \approx 2.3\times 10^{55} \;\rm erg$. The exceptionally high flux during 
this period allows the derivation of a very detailed lightcurve, as seen from Figure~1 in \cite{fermi11_3C}. The nonthermal activity lasted
for $t_{\rm full}\sim 80$~days. The onset of the activity period was characterized by a plateau stage. During the first $t_{\rm pl}\sim
13$~days, a rather steady flux was detected, with an apparent luminosity $L_{\rm
pl}\approx 10^{49}$~erg~s$^{-1}$.
The plateau stage was followed by an exceptionally bright flare, the total duration which was $t_{\rm fl}\sim 5$~days, with a
rise time of $t_{\rm r}\sim 4.5$~h. The maximum flux reached was $7\times 10^{-5} \mbox{ ph cm}^2 \mbox{s}^{-1}$, which corresponds to a
luminosity of $L_\gamma\simeq 2\times 10^{50}\rm erg\,s^{-1}$. The final stage of the flare phase was characterized by variable emission
with a flux approximately a factor of $\sim 5$ weaker than the main flare, but still a factor of $\sim 2$ above the plateau level.

The observed luminosity of the plateau phase allow us to determine a lower limit on the star mass-loss rate, which can be
derived by differentiating equation~(\ref{eq_toten}):
\begin{equation}
\dot{M}_{*}\approx 10^{23} L_{\rm pl,49} \xi^{-1}\Gamma_{0,1.5}^{-3}\,\mbox{g s}^{-1}\,,
\label{mdotw}
\end{equation}
where $L_{\rm pl,49}=L_{\rm pl}/10^{49}\,{\rm erg~s}^{-1}$.

To evaluate the feasibility of the JRGI scenario for the \c454 main flare, 
it is necessary to check whether  
the flux, the total energy release, and the flare delay with respect to the onset of the plateau,
are well described by equations~\ref{eq_delay}, \ref{eq_lum}, and \ref{eq_toten} 
for a reasonable choice of jet/star properties.

A total energy budget of the flare of $\sim 2\times10^{55}\rm erg$ is feasible, according to
equation~(\ref{eq_toten}), if
\begin{equation}
M_{\rm c,30}\Gamma_{0,1.5}^3\approx {0.04E_{\rm \gamma,55}\over \xi\bar{F}_{\rm e}}\approx {0.1\over \xi\bar{F}_{\rm
e}}\,,
\label{mass_condtion}
\end{equation}
where $M_{\rm c,30}=M_{\rm c}/10^{30}$~g is the mass of the blown up RG envelop (i.e. the initially formed cloud). 
This requirement appears to be very close to the one provided by equation~\ref{dms}:
\begin{equation}
M_{\rm c,max}\approx {5\times10^{29}\over F_{\rm e,max}}R_{*,2}^4 M_{*,0}^{-1}L_{\gamma,50}\Gamma_{0,1.5}^{-2}S_{\rm b,32}^{-1}\, \rm
g\,,
\end{equation}
where $R_{*,2}=R_*/10^2R_\sun$ and $M_{*,0}=M_*/M_\sun$, respectively.

The second term in equation~\ref{eq_delay} is expected to be short compared to the
duration of the plateau phase, even for $D\sim0.1$, and thus the duration of the 
initial plateau phase constrains the expansion time (see equation~(\ref{eq_expansion})):
$$
t_{\rm exp}\approx5.4\times10^6 F_{\rm e,max}^{1/2} \xi^{1/2}\times
$$
\begin{equation}
M_{\rm c,30}^{1/2} R_{*,2}^{-1/2} L_{\gamma,50}^{-1/2}\Gamma_{0,1.5}S_{\rm b,32}^{1/2} \rm \, s\,.
\label{eq_ex_time_2}
\end{equation}

\cite{fermi11_3C} found that the emission of the main flare consisted of 5 components (see Figure~2 in
that work): a nearly steady contribution, like a smooth continuation of the plateau emission, and 4 sub-flares of similar duration
and energetics.  In the framework of the JRGI scenario, such a description is very natural. 
The steady component would be
attributed to light clouds, continuously ejected by the star, and the four sub-flares would correspond to much heavier blobs
formed out of the blown-up stellar envelop during the initial stage.  On the other hand, the decomposition of the main flare
in four sub-flares implies a strict limitation on the variability timescale. The flare rise/decay timescales should be longer
than the blob light crossing time corrected for the Doppler boosting. 
Since the shortest variability scale was $\sim 5{\rm h}/(1+z_{\rm rs})\sim10^4$~s, 
the maximum possible size of the emitting blobs can be estimated as:
\begin{equation} 
r_{\rm b}\approx 10^{16}\Gamma_{\rm 0,1.5}\rm\,cm\,. \label{eq_cloud_size_del} 
\end{equation} 

If the jet is magnetically driven, this size constraint can be expressed through the mass of the central BH:
\begin{equation} 
{r_{\rm b}\over\omega} < 0.5\,M_{\rm BH,9}^{-1}\,, \label{eq_cloud_size} 
\end{equation} 
which is restrictive only in the case of $M_{\rm BH,9}\gg 1$. For $M_{\rm BH,9}\lesssim 1$, the blobs can cover the entire jet 
without violating the causality constraint.

In summary, if the flare detected with \fer was produced by an RG entering into the jet, the jet properties should satisfy  
to the restrictions imposed by the flux level, total energy release, and the duration of the plateau stage, respectively. Interestingly, this set of
equations allows the derivation of a {\it unique} solution, which can constrain all the key parameters through the value of
the $D$ parameter:

\begin{equation}
P_0=3\times10^6\frac{F_{\rm e,max}^{1.5}D^{1.5}}{\bar{F}_{\rm e}^{2.5}\xi z_{0,17}^{1.5}}\,\rm erg\,cm^{-3}\,,
\label{eq_lum_final}
\end{equation}

\begin{equation}
M_{\rm c}=4M_{\rm b}=5\times10^{30}\frac{F_{\rm e,max}^{1.5}{D}^{1.5}}{\bar{F}_{\rm e}^{2.5}\xi z_{0,17}^{1.5}}\,{\rm g}\,,
\label{eq_mass_final}
\end{equation}

\begin{equation}
\Gamma_0=8\left( \frac{\bar{F}_{\rm e}z_{0,17}}{F_{\rm e,max}D}\right)^{0.5}\,,
\label{eq_gamma_final}
\end{equation}
and
\begin{equation}
S_{\rm b}=8\times10^{30}\frac{z_{0,17}^{0.5}\bar{F}_{\rm e}^{1.5}}{F_{\rm e,max}^{1.5}D^{0.5}}\,{\rm cm}^2.
\label{eq_size_final}
\end{equation}
The lower limit on the jet luminosity is
\begin{equation}
L_{\rm j}>cS_{\rm b}P_0=8\times10^{47}z_{0,17}^{-1}\xi^{-1}{D\over \bar{F}_{\rm e}}\rm erg\,s^{-1}\,,
\label{eq_lum_lower_limit}
\end{equation}
which exceeds the Eddington limit for the mass of the central BH $M_{\rm BH}\sim 5\times10^8 M_{\odot}$.
To assess the feasibility of such a strongly super-Eddington jet remains out of the scope of this paper, although we note that 
\cite{lp12} have presented observational evidence indicating that such jets may not be uncommon. 

The coherent picture emanating from the jet properties derived above
suggests that the JRGI scenario can be responsible for the flare
detected from \c454 for a solution of the problem with a reasonable set of model
parameters. This solution is designed to satisfy the
requirements for (i) the total energy; (ii) the peak luminosity; and (iii) the
duration of the plateau phase. Therefore, some additional observational
tests are required to prove the feasibility of the suggested
scenario. 

Finally, the flare raise time, which is related to the blob acceleration timescale 
(see equation~\ref{dtpeak}), can be calculated for the obtained solution. Interestingly, in the limit of small $D$-values, this 
timescale appears to be independent on $D$, the only remaining free parameter, and matches closely the detected raise time of 
$t_{\rm r}\sim4.5\rm h$:
\begin{equation}
t_{\rm acc}/\left(2\Gamma_{\rm b}^2\right)\simeq5{\rm h}\,.
\end{equation}
This agreement can be treated as a cross-check that shows the feasibility of the proposed scenario.

\section{Modeling the Lightcurve and the Spectrum}
\label{sedlc}

To check whether JRGI plus synchrotron radiation can explain the
observations in the case of magnetically dominated jets (i.e., $k=1$), we have computed the lightcurve of the November 2010
flare and the SED for one of its subflares.  The radiation output was
assumed to be dominated by proton synchrotron, being external or
synchrotron self-Compton neglected due to the strong magnetic field.

\begin{figure}
\includegraphics[width=0.43\textwidth,angle=0]{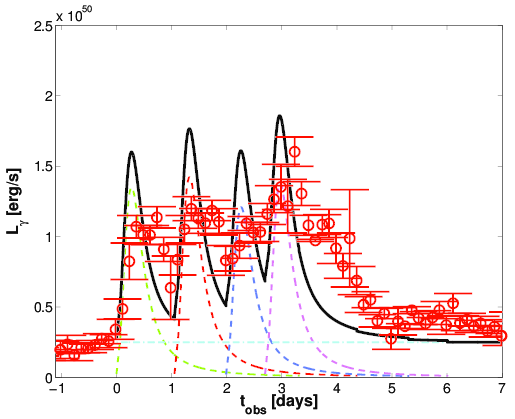}
\caption{Lightcurve computed adopting the parameters $L_{\rm j}=2.3\times 10^{48}$~erg~s$^{-1}$, $z=1.33\times 10^{17}$~cm 
$\Gamma_{\rm j}=28$, $M_{\rm c}=1.3\times10^{30}$~g, $r_{\rm c}=2.7\times 10^{15}$~cm, and $\xi=0.3$. 
We show 4 subflares  (dashed lines), plateau background
(dot-dashed line), and the sum of all of them (solid line). The observational data points and error bars are obtained from the 
\fer 3h binned count rates and photon index using luminosity distance of $D_{\rm L}=5.5\rm Gpc$ and assuming a pure powerlaw 
spectrum between 0.1 and 5~GeV.}
\label{lc}
\end{figure}

To derive the lightcurve, equation~\ref{eq_lum} has been used. In Figure~\ref{lc}, a
computed lightcurve that approximately mimics the November 2010 flare
is presented. The lightcurve has been obtained assuming four subflares
of total (apparent) energy of $10^{55}$~erg each, plus a plateau
component with luminosity of $2\times 10^{49}$~erg~s$^{-1}$. For
each subflare, we have adopted $D=0.1$. The normalization of
the lightcurve has been determined adopting the following values: the
Lorentz factor $\Gamma_0=28$, the ram pressure $P_{\rm j}=
3\times 10^6$~erg~cm$^{-3}$, blob radius $r_{\rm b}=2.7\times
10^{15}$~cm and $\xi=0.3$.  These parameters imply a minimum
jet luminosity of $L_{\rm j}=2.3\times 10^{48}$~erg~s$^{-1}$. The
remaining parameters for the emitter are $z_0=1.3\times 10^{17}$~cm
 and $M_{\rm b}=1.3\times10^{30}$~g. The
corresponding mass of the matter lost by the RG to explain the four
subflares is $5\times 10^{30}$~g, not far from the upper-limit given in
equation~\ref{dms}.

To calculate the SED, we have adopted a spectrum for the injected
protons $Q\propto E^{-p}\exp(-E/E_{\rm cut})$, and an homogeneous
(one-zone) emitter moving towards the observer with Lorentz factor
$\Gamma_{\rm b}=12$. The minimum proton Lorentz factor has been taken
equal to the shock Lorentz factor in the blob frame, i.e.  $E_{\rm
  min}=\Gamma_{\rm 0}/\Gamma_{\rm b} m_{\rm p}c^2$. The cutoff energy,
$E_{\rm cut}$, has been obtained fixing $\eta=4\times 10^3$, i.e., a relatively modest
acceleration efficiency. For the maximum proton energy, i.e. how far
beyond the cutoff the proton energy is considered, we adopted two
values: $E_{\rm max}=\infty$ and $E_{\rm max}=3E_{\rm cut}$. Regarding
the latter case, we note that assuming a sharp
high-energy cut is very natural. The injection spectrum was selected
to be hard, $p=1$, to optimize the required energetics.

\begin{figure}
\vspace{0.6cm}
\includegraphics[width=0.43\textwidth,angle=0]{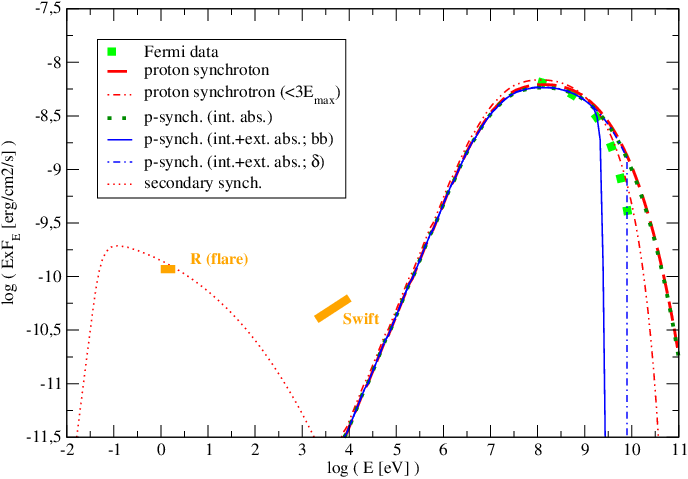}
\vspace{0.5cm}
\caption{Computed SED of the synchrotron emission for a subflare of the 2010 November. The thick dashed line shows intrinsic gamma-ray emission for the case of $E_{\rm max}=\infty$. Dotted and dot-dot-dashed line shows gamma-ray spectra corrected for internal absorption only for $E_{\rm max}=\infty$ and $E_{\rm max}=3E_{\rm cut}$, respectively.  The
thin solid and the dot-dashed lines correspond to the cases when absorption is dominated by a black body and a monoenergetic 
photon field, respectively.
The computed synchrotron SED of the secondary pairs produced via internal pair creation is also shown (dotted line).
The parameters of the flare are the same as in Figure~\ref{lc}. 
The shown observational data are from \fer, Swift 
\cite{fermi11_3C}, and the flux in the R band \cite{jor12}.}
\label{sed}
\end{figure}

In Figure~\ref{sed}, the SED of a subflare is shown. The impact of the
internal absorption on the gamma-ray spectrum is negligible, although
the emission of the secondary pairs appears in the energy band
constrained by optical measurements \cite{jor12}. For the chosen
model parameters, the synchrotron secondary component goes right
through the optical observational constraints, and for slightly higher
$z_0$-values, the secondary emission will be well below the optical
points.  Also, we note that the obtained spectrum does not violate the
X-ray upper-limits obtained by {\it Swift}.

To illustrate the impact of external $\gamma\gamma$ absorption, we
have introduced a photon field peaking at 40~eV with a luminosity
$4\times 10^{46}$~erg~s$^{-1}$, produced in a ring with radius
$10^{18}$~cm at $z=0$ around the jet base. Two photon fields have been
adopted, a black body and one represented by a $\delta$-function, to
simulate the impact of a dominant spectroscopic line.  As seen in
Figure~\ref{sed}, the impact of such an external field can be
important. The treatment of the secondary emission of the produced
pairs is beyond the scope of this work.

In addition to optical photons, radio emission was also detected at
the flare epoch and thought to be linked to the gamma-ray activity
\cite{jor12}. This radiation is strongly sensitive to the details of
the flow dynamics, and at this stage we will not try to interpret
radio observations. However, we note that the energetics involved in
gamma-ray production is very large, and JRGI comprehends complex
magnetohydrodynamical and radiative processes, so it could easily
accommodate the presence of a population of radio-emitting electrons
at $z\ge z_{\rm flare}$.

{\it Swift} X-rays could be also linked to the JRGI activity. 
X-rays may come from secondary pairs produced via
pair creation, or from a primary population of
electrons(/positrons). However, as with radio data, given the
complexity of the problem we have not tried at this stage to explain
the X-ray emission contemporaneous to the GeV flare.

\section{Discussion}

The observations of \c454 with {\it Fermi} revealed several quite
puzzling features, in particular the peculiar lightcurve, with a
nearly steady plateau phase that was interrupted by an exceptionally
bright flare. The detected flux corresponds to an apparent luminosity
of $2\times10^{50}\rm \, erg s^{-1}$, which almost unavoidably implies
a presence of a very powerful jet \cite[see e.g.][]{bgf11,kbbad13}. In the
case of powerful jets, the JRGI scenario should proceed in a quite
specific way as compared to other cases already considered in the
literature \cite{babkk10}. In particular, the mass of the material
initially removed from the star might be very large, resulting in
rather long cloud expansion and acceleration timescales, the main
flare being significantly delayed with respect to the moment of the
star entrance into the jet. The plateau emission would otherwise start
just after the jet penetration, and come from the jet crushing of
lighter clouds ejected from the stellar surface while the star travels
through the jet. The duration of the plateau phase would be determined
by the time required by the main cloud to expand and accelerate.  

We have studied the lightcurve obtained with {\it Fermi} in the context
of the JRGI scenario aiming to satisfy three main properties of the flare: total
energy, maximum luminosity and duration of the plateau stage. It was shown that
the key properties of the jet, i.e. the jet ram pressure (linked to its luminosity) and Lorentz
factor, as well as the cloud/blob characteristics, i.e. mass and cross section, can be
reconstructed as functions of the dimensionless parameter $D$. It
was also shown that in the limit of small $D$-values, the parameter space
is less demanding concerning the jet luminosity, and the key characteristics of the model
saturate at values independent of $D$, which allows conclusive cross-checks of
the scenario. In particular, the flare raise time appeared to be an independent parameter, with its value of $5\,$h closely matching the
rising time of of $4.5\,$h obtained observationally.  Furthermore, it was shown (see \cite{kbbad13}) that for the inferred
jet properties the jet-induced stellar wind can provide a mass-loss rate large enough to
generate a steady emission component with a luminosity comparable to that of the
plateau.

Although the analysis of different radiation channels involves
additional assumptions regarding the spectrum of the nonthermal
particles and density of the target fields, it was possible to show
that for magnetic fields not far below equipartition (as expected in a
magnetically launched jet) all the conventional radiation channels can be discarded, and  the emission
detected with {\it Fermi} can be produced through proton synchrotron
emission (unless $\eta \rightarrow 1$, making electron synchrotron also feasible). We note that in this case the emission
from pairs created within the blob may also explain the reported optical enhancement at the flare epoch.

Since the duration of the expansion phase determines the delay
between the onset of the plateau phase and the flare itself, it is
important to check whether the suggested scenario is consistent with
other flares registered with {\it Fermi} from the source, e.g., in
December 2009 and April 2010 \cite{agile10,fermi10}.  This issue can
be addressed through a simple scaling that relates the duration of the
plateau phase to the total energy released during the active phase:
$t_{\rm pl}\propto E_{\rm tot}^{1/2}$. Therefore, for
the previous events, with energy releases 1-2 orders of magnitude smaller
than that of the November 2010 flare,  a
rough estimation of the plateau duration 
gives plateau durations between $1.3$ and $4$ days, consistent with observations.

\bigskip 
\begin{acknowledgments}
The authors are thankful to S. Kelner for useful discussions and Benoit Lott for kindly provided observational data.  
BMV acknowledge partial  support  by  JSPS (Japan Society for the Promotion of Science):
No.2503786, 25610056, 26287056, 26800159. BMV also  MEXT (Ministry of Education, Culture, Sports, Science and Technology):
No.26105521 and RFBR  grant  12-02-01336-a.
V.B.-R. acknowledges support by the Spanish 
Ministerio de Ciencia e Innovaci\'on
(MICINN) under grants AYA2010-21782-C03-01 and FPA2010-22056-C06-02. 
V.B.-R. acknowledges financial support
from MINECO through a Ram\'on y Cajal fellowship.
This research has been supported by the Marie Curie Career Integration
Grant 321520.
\end{acknowledgments}

\bigskip 


\end{document}